\documentclass[12pt]{amsart}
\usepackage{amsmath}
\usepackage{amssymb}
\usepackage[latin1]{inputenc}
\usepackage[T1]{fontenc}
\newtheorem{lemma}{Lemma}

\newtheorem{remark}{Remark}
\newtheorem{definition}{Definition}

\newtheorem{proposition}{Proposition}

\newcommand{\bo}{{\mathcal B}([0,2\pi))}

\newcommand{\bra}{\langle}
\newcommand{\ket}{\rangle}
\newcommand{\vp}{\varphi}

\newcommand{\C}{\mathbb{C}}

\newcommand{\N}{\mathbb{N}}

\newcommand{\Z}{\mathbb{Z}}
\newcommand{\be}{\begin{equation}}
\newcommand{\eeq}{\end{equation}}
\newcommand{\bet}{\begin{equation*}}
\newcommand{\eeqt}{\end{equation*}}
\newcommand{\bea}{\begin{eqnarray}}
\newcommand{\eeqa}{\end{eqnarray}}
\newcommand{\beat}{\begin{eqnarray*}}
\newcommand{\eeqat}{\end{eqnarray*}}

\newcommand{\h}[1]{\mathcal{#1}}
\newcommand{\hil}{\mathcal{H}}

\newcommand{\hA}{\mathcal{A}}
\newcommand{\hB}{\mathcal{B}}
\newcommand{\hV}{\mathcal{V}}

\newcommand{\hF}{\mathcal{F}}

\begin{document}
\title{On infinite matrices, Schur products, and operator measures}
\author{J. Kiukas}
\address{Jukka Kiukas,
Department of Physics, University of Turku,
FIN-20014 Turku, Finland}
\email{jukka.kiukas@utu.fi}
\author{P. Lahti}
\address{Pekka Lahti,
Department of Physics, University of Turku,
FIN-20014 Turku, Finland}
\email{pekka.lahti@utu.fi}
\author{J.-P. Pellonpää}
\address{Juha-Pekka Pellonpää,
Department of Physics, University of Turku,
FIN-20014 Turku, Finland}
\email{juhpello@utu.fi}

\begin{abstract}
Measures with values in the set of sesquilinear forms on a subspace of a Hilbert space are of interest in quantum mechanics,
since they can be interpreted as observables with only a restricted set of possible measurement preparations. In this paper,
we consider the question under which conditions such a measure extends to an operator valued measure, in the concrete
setting where the measure is defined on the Borel sets of the interval $[0,2\pi)$ and is covariant with respect to shifts.
In this case, the measure is characterized with a single infinite matrix, and it turns out that a basic sufficient condition for the extensibility
is that the matrix be a Schur multiplier. Accordingly, we also study the connection between the extensibility problem and the theory of
Schur multipliers. In particular, we define some new norms for Schur multipliers.
\end{abstract}

\maketitle

\section{Introduction}
The Schur product of  infinite matrices is  an essential structure of certain quantum observables \cite{Lahti}. 
Indeed, let $X\in\mathcal B([0,2\pi))$ be a Borel subset of the interval $[0,2\pi)$ and let
\begin{equation}\label{intX}
i(X)_{nm}=\frac 1{2\pi}\int_Xe^{i(n-m)\theta}\,d\theta
\end{equation}
for each $n,m\in\Z$, so that the numbers $i(X)_{nm}$ constitute an infinite matrix $i(X)$. Let $C=(c_{nm})_{n,m\in\Z}$ be another infinite
complex matrix. Then, by definition, the Schur product of  $C$ and $i(X)$ is the matrix with entries $c_{nm}i(X)_{nm}$.
Let $\hil$ be a Hilbert space and fix an orthonormal
basis $(\vp_n)_{n\in\Z}$ of it. 
Let  $|\vp_n\ket\bra\vp_m|$
denote the operator $\vp\mapsto \bra \vp_m|\vp\ket \vp_n$, where $\bra \cdot |\cdot\ket$ 
stands for  the inner product of  $\hil$ (conjugate linear in the first argument). 
It is well-known that the formal expression
\begin{equation}\label{muoto}
E^C(X)=\sum_{n,m\in\Z} c_{nm}i(X)_{nm}\,|\vp_n\ket\bra\vp_m|
\end{equation}
constitutes, in the weak sense,  a positive normalized operator measure $X\mapsto E^C(X)$ whenever the matrix $C$ is positive 
semidefinite with diagonal elements equal to one, see e.g.
\cite{Holevo,PellonpääII}.
These operator measures describe the translation  covariant localization observables of a quantum object confined  to move in the
interval $[0,2\pi)$. If, instead of the set of integers  $\Z$ one uses the  set of natural numbers $\N$ as the index set,
  the operator measures represent the phase shift  covariant phase observables. For further details of these physical
interpretations, see e.g. \cite{PellonpääII} and the references therein. 

Although some of the properties of the studied operator measures depend crucially on the index set, the questions investigated here do not.
Namely, in the case where the index set is $J\subset \Z$, we can consider matrices $(c_{nm})_{n,m\in\Z}$ with null entries outside $J$.
This is possible because we do not, at any point, require that any matrix element is non-zero.
Hence, we take $\Z$ as the index set in this paper. 

The notion of a generalized operator measure was introduced in \cite{Pellonpää} in order to describe measurement situations where only a restricted set of state preparations
is available.
We shall see that 
the form (\ref{muoto}) defines always, that is, for any matrix $C$, a generalized operator measure, 
and the question then arises
under which conditions  it actually defines, or rather extends to, an operator measure. This question forms the subject of this paper.

\section{Basic definitions}\label{sec3}

\subsection{Preliminaries}
Let $\hV$ be a complex vector space with a countable algebraic (Hamel) basis $(\vp_n)_{n\in\Z}$.
We keep the basis fixed throughout the article. 
Let $SL(\hV)$ denote the set of complex sesquilinear forms on $\h V\times \hV$ (conjugate linear with respect to the first argument). Using the fixed 
basis,
we identify $SL(\hV)$ with the vector space of infinite complex matrices indexed by $\Z$, i.e., for any $C\in SL(\hV)$ we define a matrix $(c_{nm})_{n,m\in\Z}$ by $c_{nm}=C(\vp_n,\vp_m)$. 
We denote briefly $C\equiv(c_{nm})$. 
Let $\hV^\times$ be the algebraic antidual of $\hV$ consisting of the conjugate linear mappings $\hV\to\C$. 
Again, we indentify a $v\in\hV^\times$ with the sequence $(v_n)_{n\in\Z}$ of complex numbers given by $v_n=v(\vp_n)$ and write formally 
$(w|v)={v(w)}$, $(v|w)=\overline{v(w)}$, and $\sum_{n\in\Z}(\vp_n|v) \vp_n=v$ which we understand as $(w|v)=v(w)=\sum_{n\in\Z}\overline{w_n}v(\vp_n)$ for all $w=\sum_{n\in\Z}w_n\vp_n\in\hV$. 

Let $\hil_p$ be a completion of $\hV$ with respect to the norm $\|\,\cdot\,\|_p$ defined as
$$
\hV\ni\sum_{n\in\Z}v_n\vp_n\mapsto\left(\sum_{n\in\Z}|v_n|^p\right)^{1/p}\in[0,\infty)
$$
when $p\ge1$ and $\|\sum v_n\vp_n\|_{\infty}=\sup_{n\in\Z}|v_n|$ when $p=\infty$.
We note that $$\h V\subset\hil_p\subset\hil_2\subset\hil_q\subset\hil_\infty\subset\hV^\times$$ for all $p\in[1,2)$ and $q\in(2,\infty)$.
For $p=2$ we  write $\hil$ and $\|\,\cdot\,\|$  for $\hil_2$ and $\|\,\cdot\,\|_2$, respectively. Then $\hil$ is a Hilbert space,
the inner product of which we will denote by
$\langle\,\cdot\,|\,\cdot\,\rangle$, and we observe  that 
 $(\vp_n)_{n\in\Z}$ is   an orthonormal basis of $\hil$. 

Let $L(\hil_p,\hil_q)$ denote the Banach space of (bounded) operators  $\hil_p\to\hil_q$
equipped with the $(p,q)$-operator norm  
\begin{equation}
\| C \|_{p,q}= \sup\left\{\Big(\sum_n\Big|\sum_m c_{nm}v_m\Big|^q\Big)^{1/q} \,\Big|\, v\in\hil_p\,, \|v\|_p\leq 1 \right\}.  
\end{equation}
The $(p,q)$-norm is increasing in $p$ and decreasing in $q$, and $\| (c_{nm}) \|_{p,q}\leq\|(|c_{nm}|)\|_{p,q}$ (see, e.g.\ \cite{Bennett}).
For $p=q=2$ we write
$L(\hil)=L(\hil_2,\hil_2)$  and $\|\,\cdot\,\| =\|\,\cdot\,\|_{2,2}$.

If a matrix $C$ is such that the norm $\| C \|_{p,q}$ is finite, we say that the matrix $C$ is  $(p,q)$-\emph{bounded}.
In that case the corresponding sesquilinear form $C\in SL(\h V)$ extends to $\hil_{q^*}\times\hil_p$, 
with $q^*=q/(q-1)$,
and has the representation
\be\label{formrep}
C(u,v)= \sum_{n\in\Z}\sum_{m\in\Z} \overline{u_n}c_{nm}v_m, \ u\in \hil_{q^*},v\in\hil_p,
\eeq
where the double sum converges in either order (to the same number).
Note that for any $q\in[1,\infty)$, the topological antidual of $\hil_q$ is isomorphic to $\hil_{q^*}$.

\subsection{Schur product and Schur multiplier}\label{schur}
The Schur product of any two matrices $C=(c_{nm})_{n,m\in\Z}$ and $A=(a_{nm})_{n,m\in \Z}$ is the entrywise product $C*A=(c_{nm}a_{nm})_{n,m\in\Z}$. Any
matrix $C$ therefore defines a (Schur product) map $A\mapsto C*A$. 
In the case $1\leq p\leq q\leq\infty$, the $(p,q)$-norm of the matrix $C*A$ satisfies the following inequality \cite[Proposition 2.1]{Bennett}: 
\begin{equation}\label{Bennett}
\|C*A\|_{p,q}\leq \|C\|_{p,\infty}\,\|A\|_{1,q}.
\end{equation}

A matrix $C$ is called {\em a Schur multiplier} if $C*A$ is a $(2,2)$-bounded matrix whenever $A$ is a $(2,2)$-bounded matrix.
Therefore, Schur multipliers define linear maps in 
$L(\hil)$.  Each 
Schur multiplier map $L(\hil)\ni A\mapsto C*A\in L(\hil)$ is bounded
\cite[Theorem 1.1.1]{Larsen}.

From (\ref{Bennett}) it follows, in particular,
that any $(2,2)$-bounded matrix is a Schur multiplier.
An important (albeit trivial) example of a Schur multiplier is the matrix ${\bf 1}$ with all entries equal to 1.
The corresponding Schur multiplier map is the identity map $A\mapsto A$ in $L(\hil)$. Clearly, ${\bf 1}$ is not a $(2,2)$-bounded matrix, so a Schur multiplier
map need not arise from a $(2,2)$-bounded matrix.

Several characterizations of Schur multipliers are known. For our purposes it suffices to quote the following:  
A matrix $C$ is  a Schur multiplier
if and only if there are (norm) bounded sequences of vectors $(\psi_n)$ and $(\eta_n)$ such that $c_{nm}=\bra\psi_n|\eta_m\ket$ for all $n,m\in\Z$ \cite[Corollary 8.8]{Paulsen}.

\section{Operator measures and generalized operator measures}\label{opsec}

In this section, we define operator measures and generalized operator measures, and investigate briefly the relationship between them, in a
general level. It should be mentioned that although we have defined various spaces $\hil_p$ for the use in the subsequent chapters, the
concept of (generalized) operator measure is essentially related to the Hilbert space context. Accordingly, in this subsection, we need only to consider
$\h V$ as a dense subspace of the Hilbert space $\hil$. Also, the boundedness of a form in $SL(\h V)$ is to be understood
with respect to the Hilbert space norm.

\begin{definition}\label{defmuoto}
Let $\Omega$ be a (nonempty) set and  $\hA$ a $\sigma$-algebra of subsets of  $\Omega$.

\begin{itemize}
\item[(a)] A mapping $E:\hA\to L(\hil)$ is an \emph{operator measure}, if
the set function
\bet
\hA\ni X\mapsto \bra \vp|E(X)\psi\ket =: E_{\vp,\psi}(X)\in \C
\eeqt
is a complex measure for all $\vp,\psi\in\hil$.
The operator measure $E$ is \emph{positive} if the operators $E(X)$ are positive for all $X\in\hA$,
and {\em normalized} if $E(\Omega)=I$.

\item[(b)] A mapping $G:\hA\to SL(\hV)$ is a \emph{generalized operator measure} if the set function
\bet
\hA\ni X\mapsto G(X)(\vp,\psi)=:G_{\vp,\psi}(X)\in \C
\eeqt
is a complex measure for all $\vp,\psi\in \hV$. The generalized operator measure $G$ is {\em positive} if all the measures $G_{\vp,\vp}$, $\vp\in\hV$, 
are positive, and {\em normalized} if $G_{\vp,\psi}(\Omega)=\langle\vp|\psi\rangle$ for each $\vp,\psi\in\hV$.

\item[(c)] Let $G:\hA\to SL(\hV)$ be a generalized operator measure. If there is an operator measure $E:\hA\to L(\hil)$, such that
$\bra \vp |E(X) \psi\ket = G(X)(\vp,\psi)$ for all $X\in \hA$ and $\vp,\psi\in \h V$, we say that $G$ is \emph{extensible}.
Since $\hV$ is dense, the extension of $G$, when exists, is unique. We denote it by $\overline G$.

\end{itemize}
\end{definition}

The set of generalized operator measures
$G:\hA\to SL(\hV)$ forms a linear space (with respect to the pointwise operations) and 
the extensible ones constitute a subspace of this vector space.

Let $\hV_1=\{ \vp\in \h V\mid \|\vp\|\leq 1\}$.
For each generalized operator measure $G:\hA\to SL(\hV)$, we denote
\bet
\|G\| = \sup\{ |G_{\vp,\psi}(X)|\mid \vp,\psi\in\h V_1, X\in \h A\}.
\eeqt
By using the following proposition it is easy to see that the map $G\mapsto \|G\|$ is actually a Banach space norm when restricted to the subspace of extensible generalized operator measures. We will consider this norm in Section \ref{normsec}, in the concrete case relevant to this paper.

The following proposition gives a general characterization for the extensibility of a generalized operator measure $G$ in terms of the boundedness of the forms $G(X)$.

\begin{proposition}\label{prop1}
Let $G:\hA\to SL(\hV)$ be a generalized operator measure.
\begin{itemize}
\item[(a)] The following conditions are equivalent.
\begin{itemize}
\item[(i)] $G$ is extensible;
\item[(ii)] $\|G\| <\infty$;
\item[(iii)] $G(X)$ is bounded for each $X\in \h A$.
\end{itemize}
\item[(b)] If $\h F\subset \h A$ is an algebra which generates the $\sigma$-algebra $\h A$, then
the condition
\begin{itemize}
\item[(ii')] $\sup\{ |G_{\vp,\psi}(X)|\mid \vp,\psi\in\h V_1, X\in \h F\}<\infty.$
\end{itemize}
is equivalent to each of the conditions of (a).
\end{itemize}

\end{proposition}
\begin{proof} Let $\h F$ be as in part (b) of the proposition. (It suffices to prove (b), for then (a) follows by putting $\h F=\h A$.)
First we note that (i) implies (ii), because the range of any operator
measure is norm bounded (as a consequence of the uniform boundedness theorem and the fact that the range of each complex measure
is bounded). Now (ii) implies (ii') trivially.

To prove that (ii') implies (iii), let $M$ be the supremum in (ii'), so the assumption is that $M <\infty$. Denote
\bet
\hB = \{ X\in\hA\mid |G_{\vp,\psi}(X)|\leq M \text{ for all } \vp,\psi\in \h V_1\}.
\eeqt
Then, by assumption, $\hF\subset \hB$. We prove that $\hB$ is a monotone class.
First, let $(X_n)$ be an increasing sequence of sets in $\h B$, and let $\vp,\psi\in \h V_1$. Then we have
\bet
G_{\vp,\psi }\Big(\bigcup_{n=1}^\infty X_n\Big)= \lim_{n\to\infty} G_{\vp,\psi }(X_n),
\eeqt
because $X\mapsto G_{\vp,\psi }(X)$ is a complex measure. Since $|G_{\vp,\psi }(X_n)|\leq M$ for all $n\in \N$,
also $|G_{\vp,\psi }(\bigcup_{n=1}^\infty X_n)|\leq M$, so $\bigcup_{n=1}^\infty X_n\in \hB$.
Similarly, for a decreasing sequence $(X_n)$ in $\hB$, we get $\bigcap_{n=1}^\infty X_n\in \hB$. Hence, $\hB$ is a monotone class.
It now follows from the monotone class
theorem, that the monotone class generated by $\hF$ is $\hA$. But $\hF\subset \hB$, so $\hA=\hB$, which
means that $G(X)$ is bounded for all $X\in \h A$. Hence, (iii) holds.

It remains to prove that (iii) implies (i).
Since $\h V$ is dense, (iii) implies that for each $X\in \hA$, there is an $E(X)\in L(\hil)$, such that
$G(X)(\vp,\psi)= \bra \vp |E(X) \psi\ket$ for all $\vp,\psi\in \h V$. Let $\vp,\psi\in\hil$, and choose
the sequences $(\vp_k)_{k\in\N}$, $(\psi_k)_{k\in\N}$ of vectors of $\h V$ converging to $\vp$ and $\psi$, respectively. By definition,
each set function $X\mapsto \bra \vp_k|E(X)\psi_k\ket =G(X)(\vp_k,\psi_k)$ is a complex measure. Since
$\lim_{k\to\infty} \bra \vp_k|E(X)\psi_k\ket = \bra \vp|E(X)\psi\ket$ for each $X\in\hA$, it follows from the
Nikod\'ym convergence theorem (see e.g. \cite[p. 160]{Dunford}) that also $X\mapsto \bra \vp|E(X)\psi\ket$ is a complex measure. Hence $X\mapsto E(X)$ is
an operator measure. The proof is complete.
\end{proof}

\begin{remark}\label{remarkprop1} {\rm It should be emphasized that although the algebras $\h F$ and $\h A$ are in similar roles in (ii) and (ii'),
the $\sigma$-algebra $\h A$ cannot, in general, be replaced by the algebra $\h F$ in (iii). This is basically a consequence of the fact that the
Nikod\'ym boundedness theorem fails in the case of finitely additive measures defined on set algebras, except in certain special cases \cite{Seever}.
In the relevant concrete case where $\h A$ is the Borel $\sigma$-algebra of the interval $[0,2\pi)$ and $\h F$ the algebra of finite unions
of subintervals, we will give an explicit counterexample which shows that $\h A$ cannot be replaced by $\h F$ in (iii)
(see Proposition \ref{thetalemma} of Section \ref{covariantsec}).
}\end{remark}

If $G$ is a generalized operator measure
and $f:\Omega\to\C$ a bounded measurable function, such that $f$ is $G_{\vp,\psi}$-integrable for all $\vp,\psi\in \h V$, then the map
\bet 
\h V\times \h V\ni (\vp,\psi)\mapsto \int_\Omega f\,dG_{\vp,\psi}\in \C
\eeqt
is a sesquilinear form,  which we denote by $\int f dG$. 
If $G$ is extensible then there is a
unique bounded linear operator $L(f,\overline G)$ such that $
\bra\vp|L(f,\overline G)\psi\ket = \int_\Omega f\,d{\overline G}_{\vp,\psi}$,
for all $\vp,\psi\in\hil$, see, e.g. \cite[Theorem 9]{Berberian}. 
Clearly, in this case one also has
$\bra\vp|L(f,{\overline G})\psi\ket=\int_\Omega f\, dG_{\vp,\psi}$,
for all $\vp,\psi\in\hV$.

\section{$\Z$-covariant operator measures on $ \hB([0,2\pi))$}\label{covariantsec}

\subsection{Definitions}
Take any infinite complex matrix $C=(c_{nm})_{n,m\in\Z}$, and  define, for each $X\in \hB([0,2\pi))$,
as a specification of (\ref{muoto}),  
\begin{equation}\label{muotoC}
G^C(X)(\vp,\psi) = \sum_{n,m\in\Z} c_{nm} i(X)_{nm}\bra\vp| \vp_n\ket\bra \vp_m|\psi\ket, \ \ \psi,\vp\in \h V.
\end{equation}
Clearly, $G^C(X)(\vp_k,\vp_l)=c_{kl}i(X)_{kl}$ for all $k,l\in\Z$, so that the matrix of $G^C(X)$ 
is  the Schur product of $C$ and $i(X)$.
Each $G^C(X)$ is a sesquilinear form on $\h V\times \h V$ and the map $X\mapsto G^C(X)$ is a generalized operator measure, 
which is  covariant with respect
to the representation 
$\theta \mapsto e^{i\theta Z} := \sum_ne^{i\theta n} |\vp_n\rangle\langle \vp_n|$.
Conversely,
any such covariant generalized operator measure  $G$
is of this form for a unique $C=(c_{nm})$ \cite{Pellonpää}. 
The matrix $C$
is  called the \emph{structure matrix} of $G$.

For each $k\in\Z$, let $V^C_k:\hV\to \hil$ 
 be a linear operator defined by $V^C_k( \vp_n) = c_{n-k,n}\vp_{n-k}$ for all $n\in \Z$.
The form $(\vp,\psi)\mapsto \bra \vp |V^C_k\psi\ket$
is easily seen to be the $k$th \emph{cyclic moment} of $G^C$, i.e.
\bet
 \bra \vp |V^C_k\psi\ket =
\int_0^{2\pi} e^{ik\theta}\,G^C_{\vp,\psi}(d\theta).  \ \ \psi,\vp\in \h V.
\eeqt

For each $s\in\N$, let $\Theta_s^C$ denote the $s$th \emph{moment} of $G^C$, 
\bet
\Theta_s^C(\vp,\psi) = \int_0^{2\pi} \theta^s G^C_{\vp,\psi}(d\theta).  \ \ \psi,\vp\in \h V.
\eeqt
Then, in particular,
\be\label{firstmoment}
\Theta_1^C(\vp_n,\vp_m)=
\begin{cases}
\frac{1}{ i(n-m)}c_{nm}, & n\neq m, \\  
\pi c_{nn},  & n=m.
\end{cases}
\eeq

Since the functions $\theta\mapsto \theta^s$ and $\theta\mapsto e^{ik\theta}$ are bounded,  the following
result is immediate.

\begin{proposition}\label{firstmomentcond} If $G^C$ is extensible, then the forms $\Theta_s^C$ are bounded, and each operator $V^C_k$ is bounded. 
\end{proposition}

We proceed to study some boundedness conditions related to the extensibility of the generalized operator measure $G^C$.
The first thing to note is that 
for each $(2,2)$-bounded matrix $C$, the form $G^C$ is extensible by Proposition \ref{prop1} (a). This is because
$\|\int_X e^{i\theta Z}Ce^{-i\theta Z}d\theta\|\le 2\pi\|C\|$. 
Obviously,
the extensibility of a $G^C$ does not require $C$ to be a $(2,2)$-bounded matrix. This is demontrated e.g. by the  important example of  $G^{\bf 1}$.
The associated operator measure $\overline{G^{\bf 1}}$
is known to be unitarily equivalent to the canonical spectral measure of $L^2([0,2\pi))$
via the unitary transformation  between the basis $\{ \vp_n\mid n\in\Z\}$ and the Fourier basis of $L^2([0,2\pi))$, see, e.g.\ \cite{PellonpääII}.
In particular, each $i(X)$ is a $(2,2)$-bounded matrix describing a projection operator.

It seems to be difficult to give a general characterization of the matrices $C$ for which $G^C$ is extensible.
Indeed, it is usually assumed that $G^C$ is positive or, equivalently, $C$ is positive semidefinite,
in which case it is extensible if and only if $\sup_{n\in\Z}c_{nn}<\infty$.
This happens exactly when there is a  bounded sequence
of vectors $(\psi_n)_{n\in \Z}$, such that $c_{nm}= \bra \psi_n|\psi_m\ket$ for all $n,m\in \Z$ \cite{PellonpääII}. 
In this case, the positive operator measure $\overline{G^C}$ is normalized   
if and only if $c_{nn}=\|\psi_n\|^2=1$ for all $n\in\Z$.
More generally, if $C$ is a Schur multiplier,
then $G^C$ is extensible (Section~\ref{Schur}).

\subsection{Boundedness of the matrix elements}
Here we consider some boundedness criteria involving the matrix elements of the structure matrix $C$.
We denote $S^C= \sup_{n,m} |c_{nm}|$.
The first result is easy to prove using the cyclic moments.
\begin{proposition}\label{propcyclic}
Assume that $G^C$ is extensible. Then $S^C <\infty$.
\end{proposition}
\begin{proof} Let $\overline{G^C}$ denote  the operator measure determined by 
$G^C$.
For each $k\in \Z$, let $f_k:[0,2\pi)\to \C$ be the function with $f_k(\theta)= e^{ik\theta}$. Now each cyclic moment form
is bounded, so each $V^C_k$ is a bounded operator, with $L(f_k,\overline{G^C})=V^C_k$.
It follows from Proposition \ref{prop1} (a) (condition (ii)) that $\|G^C\|<\infty$. Now
\beat
|\bra \psi|V^C_k\vp\ket| &=& |\bra \psi |L(f_k,\overline{G^C})\vp\ket| \leq  \int |f_k| d|G^C_{\psi,\vp}| =  \int d|G^C_{\psi,\vp}|\\
&=& |G^C_{\psi,\vp}|([0, 2\pi))\leq 4 \sup_{X\in {\bo}} |G^C_{\psi,\vp}(X)|\leq 4\|G^C\|
\eeqat
for all unit vectors $\psi,\vp\in\h V$. 
Hence, $\|V^C_k\|\leq 4\|G^C\|$ for all $k\in \Z$.
But $\|V_k^C\|=\sup \{|c_{n,n+k}|\mid n\in\Z\}$ for all $k\in \Z$, so ${\rm sup}_{n,m\in\Z}|c_{nm}|<\infty$.
\end{proof}

\begin{remark}\label{remark2} {\rm The converse of the preceding proposition does not hold. Indeed, let $c_{nm}=1$ for $n > m$,
$c_{nm}=-1$ for $n<m$, and $c_{nn}=0$ for all $n$. Then the nondiagonal matrix elements of the first moment form 
$\Theta^C_1$ are
$\Theta^C_1(\vp_n,\vp_m)= -i|n-m|^{-1}$, so $\Theta^C_1$ 
is not a $(2,2)$-bounded matrix (see e.g. \cite[p. 258]{matriisikirja}), 
and hence $G^C$ cannot be extensible by Proposition \ref{firstmomentcond}.
}\end{remark}

In the following proposition, we compare the conditions
$\|C\|_{1,\infty}<\infty$ and $\|C\|_{2,\infty}<\infty$, in view of the extensibility of $G^C$.
We recall that $\|C\|_{1,\infty}=S_C$ and $\|C\|_{2,\infty}= \sup_n \sqrt{ \sum_m |c_{nm}|^2}$.

\begin{proposition}\label{propinfty} Let $C$ be a matrix.
\begin{itemize}
\item[(a)] Assume that $\|C\|_{1,\infty}<\infty$. Then, for each bounded measurable function $f:[0,2\pi)\to \C$, the
matrix $\int f dG^C$ is an element of $L(\hil_1, \hil)$, with
\bet
\left\|\int f dG^C\right\|_{1,2} \leq S^C\sup_{\theta\in [0,2\pi)}|f(\theta)|.
\eeqt
In particular, $G^C(X)\in L(\hil_1,\hil)$ for all $X\in \bo$, with
\bet
\sup_{X\in \bo}\|G^C(X)\|_{1,2}\leq S^C.
\eeqt
\item[(b)] Assume that $\|C\|_{2,\infty}<\infty$. Then $G^C$ is extensible.
\end{itemize}
\end{proposition}
\begin{proof}
Assume that $C$ is such that $S^C<\infty$. To prove (a), let $M= \sup_{x\in [0,2\pi)}|f(x)|$. 
Since $f$ is bounded, $\int f dG^C$ exists. Clearly, $\int f dG^C= C*\int f dG^{{\bf 1}}$.
Since $\overline{G^{{\bf 1}}}$ is a spectral measure, the operator integral $L(f,\overline{G^{{\bf 1}}})$ coincides
with the usual spectral integral, so we have $\|L(f, \overline{G^{{\bf 1}}})\|\leq M$.
The inequality (\ref{Bennett}) now gives
\bet
\left\|\int f dG^C\right\|_{1,2} \leq \|C\|_{1,\infty}\left\|\int f dG^{{\bf 1}}\right\|_{1,2}\leq \|C\|_{1,\infty}\left\|\int f dG^{{\bf 1}}\right\|_{2,2}\le S^C M.
\eeqt
Since each characteristic function $\chi_X$, $X\in \bo$, is bounded and measurable, (a) is proved.

To prove (b), we use a stronger estimate from (\ref{Bennett}): 
$$
\|C*i(X)\|\le
\|C\|_{2,\infty}\| i(X)\|_{1,2}\le\|C\|_{2,\infty}\| i(X)\|
$$
for all $X\in \bo$. Hence, if $\|C\|_{2,\infty}<\infty$, then $G^C$ is extensible by Proposition \ref{prop1} (a) and by the
fact that each $i(X)$ is a projection operator.
\end{proof}

\begin{remark} {\rm As the example in Remark \ref{remark2} showed, the condition $S^C<\infty$ is not enough for the extensibility of $G^C$.
However, part (a) of the preceding proposition shows that the form $G^C$ is still ''extensible'' in a weaker sense.
The condition $\|C\|_{2,\infty}<\infty$ ensures the extensibility, but it is too strong, since it rules out
e.g. the important matrix ${\bf 1}$. Note, however, that this condition is weaker than the condition $\|C\|<\infty$.
}\end{remark}

Next we discuss some implications of boundedness of the diagonals of the matrix $C$.
 Denote $C_k = \sup \{|c_{n,n+k}|\mid n\in\Z\}$.

\begin{proposition}\label{propo5}
If $C_k<\infty$ for all $k\in\Z$ and the sequence
$\left( \sum_{k=-N}^N i(X)_{0,k} V^C_k\right)_{N\geq 0}$ converges weakly in $L(\hil)$ for all $X\in\bo$,
then $G^C$ is extensible.
\end{proposition}

\begin{proof} Since $C_k<\infty$, the cyclic moment $V_k^C$ is bounded with $\|V^C_k\|=C_k$. 
For each $X\in \bo$, let $E(X)\in L(\hil)$ be the limit of the sequence.
 Since
\bet
\bra \vp_n|E(X)\vp_m\ket = \sum_{k\in \Z} i(X)_{0,k} c_{m-k,m}\bra \vp_n|\vp_{m-k}\ket = i(X)_{0,m-n}c_{nm} = G^C(X)( \vp_n,\vp_m),
\eeqt
$G^C$ is extensible by Proposition \ref{prop1} (a). 
\end{proof}

To prove the next proposition we need the following lemma.

\begin{lemma}
Let $A\in L(\hil)$ and denote
$A_k = \sum_{n\in\Z}\bra\vp_n|A\vp_{n+k}\ket|\vp_n\ket\bra\vp_{n+k}|, \ \ k\in \Z$.
Then $A_k\in L(\hil)$ for each $k\in \Z$ and 
$A = \lim_{M\to\infty}\frac{1}{M}\sum_{N=0}^{M-1}\sum_{k=-N}^N A_k$
in the weak sense.
\end{lemma}
\begin{proof}
Let $A\in L(\hil)$ and define a covariant operator measure $\overline{G^A}$. Then, for any $k\in\Z$, we have $A_k=\int_0^{2\pi}e^{ik\theta}\overline{G^A}(d\theta)$.
Now, each $A_k$ is bounded. 
For any $M=1,2,...$,
$$
\left\|\frac{1}{M}\sum_{N=0}^{M-1}\sum_{k=-N}^N A_k\right\|=\left\|\int_0^{2\pi}K_M(\theta)\overline{G^A}(d\theta)\right\|\le\frac{1}{2\pi}\int_0^{2\pi} K_M(\theta)\|e^{i\theta Z}Ae^{-i\theta Z}\|d\theta\le\|A\|,
$$
where
$$
K_M(\theta)=\frac{1}{M}\sum_{N=0}^{M-1}\sum_{k=-N}^Ne^{ik\theta}= \frac{1}{M}\left[\frac{\sin(M\theta/2)}{\sin(\theta/2)}
\right]^2
$$
is the Fej\'er kernel \cite[p.17]{Hoffman}. Therefore,
the sequence $M\mapsto{M}^{-1}\sum_{N=0}^{M-1}\sum_{k=-N}^N A_k$
is norm bounded. Since 
$$
\lim_{M\to\infty}
\left\bra\vp_n\bigg|\frac{1}{M}\sum_{N=0}^{M-1}\sum_{k=-N}^N A_k\vp_m\right\ket=
\bra\vp_n|A\vp_m\ket\lim_{M\to\infty}\left(1-\frac{|n-m|}{M}\right)
=\bra\vp_n|A\vp_m\ket
$$
for all $n,m\in\Z$, it follows that
$$
\lim_{M\to\infty}\frac{1}{M}\sum_{N=0}^{M-1}\sum_{k=-N}^N A_k=A
$$
weakly on $\hil$.
\end{proof}

\begin{proposition}
$G^C$ is extensible if and only if $C_k<\infty$ for all $k\in\Z$ and
$$M\mapsto \frac{1}{M}\sum_{N=0}^{M-1}\sum_{k=-N}^N i(X)_{0,k} V^C_k$$ converges weakly in $L(\hil)$ for all $X\in\bo$. In this case, for any $\vp,\psi\in\hil$,
$$
\lim_{M\to\infty}
\frac{1}{M}\sum_{N=0}^{M-1}\sum_{k=-N}^N e^{-ik\theta}\bra\vp|V^C_k\psi\ket=g^C_{\vp,\psi}(\theta)
$$
with respect to the $L^1$-norm,
where $g^C_{\vp,\psi}$ is the density of the complex measure $\overline{G^C_{\vp,\psi}}$.
\end{proposition}
\begin{proof}
Let $G^C$ be extensible and fix $X\in \bo$. Applying the preceding lemma to the operator $A=\overline{G^C}(X)$, and noting that then $A_k=i(X)_{0,k} V^C_k$, we get
$$
\lim_{M\to\infty} \frac{1}{M}\sum_{N=0}^{M-1}\sum_{k=-N}^N i(X)_{0,k} V^C_k=\overline{G^C}(X)
$$
weakly for any $X\in\bo$. To prove the converse, suppose first that each $C_k$ is finite, so that each $V^C_k$ is bounded. Then assume also that
the sequence $M\mapsto\frac{1}{M}\sum_{N=0}^{M-1}\sum_{k=-N}^N i(X)_{0,k} V^C_k$ converges weakly to some bounded operator, say, $E^C(X)$. Since 
$$
\bra\vp_n|E^C(X)\vp_m\ket=\lim_{M\to\infty}
\left\bra\vp_n\bigg|\frac{1}{M}\sum_{N=0}^{M-1}\sum_{k=-N}^N i(X)_{0,k} V^C_k
\vp_m\right\ket
=G^C(X)(\vp_n,\vp_m)
$$
it follows from Proposition \ref{prop1} (a) that $G^C$ is extensible and $\overline{G^C}=E^C$.

Let $\vp,\psi\in\hil$. Suppose then that $g^C_{\vp,\psi}\in L^1([0,2\pi))$ is the density of the complex measure $\overline{G^C_{\vp,\psi}}$. Now its Fourier coefficients are of the form
$$
\frac{1}{2\pi}\int_0^{2\pi}e^{ik\theta}g^C_{\vp,\psi}(\theta)d\theta=\bra\vp|V_k^C\psi\ket
$$
and the sequence 
$$
M\mapsto\frac{1}{M}\sum_{N=0}^{M-1}\sum_{k=-N}^N e^{-ik\theta}\bra\vp|V^C_k\psi\ket
$$
of the Cecaro means
converges to $g^C_{\vp,\psi}$ with respect to the $L^1$-norm \cite[p.\ 17]{Hoffman}.
\end{proof}

To conclude this subsection, we note that the condition $S^C<\infty$ allows us to  define the relevant complex measure
$G^C_{\vp,\psi}$ also for the case where $\vp\in \hil$ and $\psi\in \hil_1$. We recall that in the case where $G^C$ is extensible, the measure can
be defined for all $\vp,\psi\in \hil$.

First of all, it follows from part (a) of Proposition \ref{propinfty} that each $\|G^C(X)\|_{1,2}$ is finite. Hence, for each $\psi\in \hil_1$, $\vp\in \hil_2$, and $X\in \h B[0,2\pi)$, the double sum
\bet
\sum_{n\in \Z} \sum_{m\in\Z} i(X)_{nm} c_{nm} \bra \vp|\vp_n\ket\bra\vp_m|\psi\ket
\eeqt
converges to some number, which does not depend on the order of summation (see \eqref{formrep}). We call that number $G^C_{\vp,\psi}(X)$, because it coincides with $G^C_{\vp,\psi}(X)$ if $\vp,\psi\in\h V$.

It follows by the Nikod\'ym convergence theorem \cite[p. 160]{Dunford} that the map $X\mapsto G^C_{\vp,\psi}(X)$ is a complex measure.
In addition, it has a density function. This follows from the Radon-Nikod\'ym theorem, because $i(X)$ is the null matrix if the set
$X$ has zero Lebesgue measure.

The density of $G^C_{\vp,\psi}$ can be explicitly calculated, and it turns out that it belongs to $L^2([0,2\pi))$. This is seen as follows.
For all $\vp\in\hil$ and $\psi\in\hil_1$, the expression
\bet
g_{\vp,\psi}^C(\theta)=\sum_{m\in\Z}\left(\sum_{n\in\Z}e^{i(n-m)\theta}c_{nm}\langle\vp|\vp_n\rangle\bra\vp_m|\psi\ket \right)
\eeqt
defines an element $g_{\vp,\psi}^C$ of $L^2([0,2\pi))$ by the following calculation. Define for all $m\in\Z$ the $L^2$-function $\Phi_m$ by
$$
\Phi_m(\theta)=e^{-im\theta}\bra\vp_m|\psi\ket \sum_{n\in\Z}c_{nm}\langle\vp|\vp_n\rangle e^{in\theta}.
$$
Clearly, $\|\Phi_m\|_{L^2}\le S^C|\bra\vp_m|\psi\ket |\|\vp\|$, so the sequence
$$
\N\ni N\mapsto \sum_{m=-N}^N \Phi_m\in L^2([0,2\pi))
$$
is Cauchy, thereby converging to an $L^2$-function. In addition, the limit is integrable. This is because
$L^2([0,2\pi))\subset L^1([0,2\pi))$ by the Cauchy-Schwarz inequality. 
Similarly, $L^2$-convergence implies $L^1$-convergence. Now we get
\beat
G^C_{\vp,\psi}(X)&=& \sum_{m\in \Z} e^{-im \theta}\bra\vp_m|\psi\ket  \left(\sum_{n\in Z}\int_X  c_{nm} \bra \vp|\vp_n\ket e^{in \theta}\right)\\
&=&\sum_{m\in \Z} \int_X \Phi_m(\theta)d\theta = \int_X \sum_{m\in \Z} \Phi_m(\theta)d\theta = \int_X g_{\vp,\psi}^C(\theta) d\theta,
\eeqat
where the second equality follows because each series $N\mapsto \sum_{n=-N}^N c_{nm} \bra \vp|\vp_n\ket e^{in \theta}$
converges in $L^2$ and hence in $L^1$, and the third equality for a similar reason.

Thus, the Radon-Nikod\'ym derivative of $G^C_{\vp,\psi}$ is indeed the function $g_{\vp,\psi}^C$.

\subsection{Boundedness of the first moment}

Since the first moment form completely determines the generalized operator measure $G^C$ (see \eqref{firstmoment}),
it is natural to ask if the boundedness of $\Theta_1^C$ already implies the extensibility of $G^C$.
The following result shows that the answer is negative, but that $G^C(X)$ is actually bounded for each
$X\in \h F$, where $\h F$ is the algebra of finite unions of subintervals of $[0,2\pi)$.

\begin{proposition}\label{thetalemma}
\begin{itemize}
\item[(a)] Assume that $\Theta_1^C$ is bounded. Then $G^C([a,b))$ is bounded for all $a,b\in [0,2\pi]$, $a\leq b$.
\item[(b)] Boundedness of $\Theta_1^C$ does \emph{not} imply the extensibility of $G^C$.
\end{itemize} 
\end{proposition}
\begin{proof} To prove (a), let $\Theta_1^C$ denote also the bounded operator associated with the form $\Theta_1^C$. It follows that also
the diagonal elements $\Theta_1^C(\vp_n,\vp_n)$ alone constitute a $(2,2)$-bounded matrix, which we
denote by  ${\rm dg}(\Theta^C_1)$.
Now, if $n\in\Z$, we have
$$
G^C([a,b))( \vp_n, \vp_n) 
=\frac {1}{2\pi^2} (b-a)\bra\vp_n |{\rm dg}(\Theta_1^C) \vp_n\ket,
$$
and, for $n\neq m$, we get
$$
G^C([a,b))( \vp_n,\vp_m) 
= \frac{1}{2\pi} \bra \vp_n| (e^{ibZ}\Theta^C_1 e^{-ibZ}-e^{iaZ}\Theta^C_1 e^{-iaZ})\vp_m\ket.
$$
Hence, the matrix elements of $G^C([a,b))$ are those of the bounded operator
\bet
\frac {1}{2\pi} \left(\frac {1}{\pi}(b-a){\rm dg}(\Theta^C_1) + e^{ibZ}\Theta^C_1 e^{-ibZ}-e^{iaZ}\Theta^C_1 e^{-iaZ}\right),
\eeqt
and so $G^C([a,b))$ is bounded. This proves (a).

To prove (b), let $B$ be any bounded operator on $\hil$ with the matrix elements $b_{nm}$.
Then, by defining a matrix $C^B$ with the elements $c^B_{nm}=i(n-m)b_{nm}+\pi^{-1}b_{nn}\delta_{nm}$, we get a generalized operator measure 
$G^{C^B}$ for which the first moment form $\Theta^{C^B}_1$ is a restriction of $B$ and, thus, bounded.
Let $\gamma=\sum_{n=1}^\infty(n^{-1}\ln n)\vp_n\in\hil$. 
Choose $B =|\gamma\rangle\langle\vp_0|$.
Then $c^{B}_{n0}=i\ln n$ for all $n\ge1$ and $\sup_{n,m\in \Z} |c^{B}_{nm}|=\infty$.
By Proposition \ref{propcyclic}, $G^{C^{B}}$ is not extensible.

\end{proof}
\begin{remark} {\rm 
It follows from part (b) of Proposition \ref{prop1}, that although the boundedness of the first
moment implies that $\|G^C(X)\|<\infty$ for each $X\in \h F$ (Proposition \ref{thetalemma} (a)), it does not imply that
$\sup_{X\in \h F} \|G^C(X)\|<\infty$. 
}\end{remark}

\begin{proposition}\label{prop6}
Assume that $\Theta_1^C$ is bounded. Then $V_k^C$ is bounded for all $k\in\Z$. The converse does not hold.
\end{proposition}
\begin{proof}
By assumption, the operator $\sum_{n\ne m}\frac{c_{nm}}{n-m}|{\vp_n}\ket\bra{\vp_m}|$ is bounded with the norm $N$, say, so it follows that
$\|V_k^C\|=\sup_{n\in\Z}|c_{n,n+k}|\le N|k|$ for all $k\in\Z\setminus\{0\}$.
The case $k=0$ follows from the boundedness of the diagonal operator of the first moment.
The fact that the converse result does not hold can be seen by choosing the matrix $C$ used in Remark \ref{remark2}.
\end{proof}

\begin{proposition}
Assume that $\Theta_1^C$ is bounded. Then $\Theta_s^C$ is bounded for all $s\in\N$
and the operator sequence $(\Theta_s^C)_{s\in\N}$ is exponentially bounded in the sense that $\|\Theta_s^C\|\le R \pi^ss!$ for all $s\in\N$ where $R>0$.
\end{proposition}

\begin{proof}
Define, for all $l=1,2,3,...,$
$$
A_l=\sum_{n\ne m\in\Z}\frac{c_{nm}}{(n-m)^l}|{\vp_n}\ket\bra{\vp_m}|,
$$
and $A_0=\sum_{n\in\Z}c_{nn}|\vp_n\ket\bra\vp_n|$.
Since, by assumption, $\Theta_1^C$ is bounded, also $A_0=-\Theta_0^C$ and $A_1$ are bounded. 
Define $B_l=\sum_{n\ne m\in\Z}\frac{1}{(n-m)^l}|{\vp_n}\ket\bra{\vp_m}|$. Now $B_1$ is bounded with $\|B_1\|=\pi$. (This follows from
the fact that $-iB_1+\pi I$ can be interpreted as the first moment of the canonical spectral measure on $L^2([0,2\pi))$.) Since
$B_{l+1}=B_1*B_l$, it follows by induction and using (\ref{Bennett}), that $B_l$ is bounded for all $l=1,2,...$.
Then it follows that $A_{l+1}=A_1*B_l$ is bounded for all $l=1,2,...$. From
\begin{eqnarray*}
\frac{1}{2\pi}\int_0^{2\pi}\theta^se^{ik\theta}d\theta=
\begin{cases}
-s!\sum_{l=1}^s\frac{i^l(2\pi)^{s-l}}{(s-l+1)!}\cdot\frac{1}{k^l}, & s\ge 1,\;k\ne 0, \\
\frac{(2\pi)^{s}}{s+1}, & s\ge1,\;k=0.
\end{cases}
\end{eqnarray*}
it follows that
$$
\Theta_s^C=-s!\sum_{l=0}^s\frac{i^l(2\pi)^{s-l}}{(s-l+1)!}A_l
$$
is bounded for all $s\in\N$. Moreover, since also $\|A_{l}\|\le\|B_1\|^{l-1}\|A_1\|=\pi^{l-1}\|A_1\|$ for all $l=1,2,...$,
$$
\|\Theta_s^C\|\le s!(2\pi)^s\sum_{l=0}^s\frac{(2\pi)^{-l}\|A_l\|}{(s-l+1)!}\le s!(2\pi)^s
\left[\frac{\|A_0\|}{(s+1)!}+\frac{\|A_1\|}{\pi}\sum_{l=1}^s\frac{1}{2^l(s-l+1)!}
\right].
$$
Since $\sum_{l=1}^s\frac{1}{2^l(s-l+1)!}\le2^{-s-1}\sum_{j=1}^\infty\frac{2^j}{j!}=2^{-s-1}(e^2-1)$, one can choose any $$R\ge\|A_0\|+(2\pi)^{-1}(e^2-1)\|A_1\|.$$
\end{proof}
\begin{remark} {\rm
We want to point out an interesting implication of the above proposition. Namely, the result allows us to define the continuous exponential mapping
\bet
{\mathcal D}\ni z\mapsto F^C(z)=\int_0^{2\pi}e^{z\theta}d G^C(\theta)\in L(\hil),
\eeqt
where $\mathcal D$ is 
the open disk in the complex plane centered at the origin and having radius $\pi^{-1}$.
Restricting $z$ to be real or pure imaginary, one gets the (bounded) Laplace or Fourier transforms of $G^C$ (defined on $(-\pi^{-1},\pi^{-1})$).
In the case of a (real) positive operator measure, one can reconstruct the measure from its Laplace or  Fourier transform, but in the case of a (generalized) operator measure it is not clear when this is possible.
} \end{remark}

\section{Schur multipliers and generalized vectors}\label{Schur}

For each Schur multiplier $C$, the generalized operator measure
$G^C$ is extensible. Indeed,
if $C$ is a Schur multiplier, then each $G^C(X)=C*i(X)$ is bounded, 
and so $G^C$ is extensible by Proposition \ref{prop1} (a). We formulate this basic observation as a proposition:
\begin{proposition}\label{basicSchur}
A  $\Z$-covariant  generalized operator measure is extensible whenever its structure matrix is a Schur multiplier.
\end{proposition}

Next we want to consider generalized operator measures in terms of \emph{generalized vectors}, i.e. the elements of
$\h V^\times$. 
For any $v\in\mathcal V^\times$, define a linear operator $\mathcal{V}\to \hV$ by
\bet
V_v=\sum_{n\in\Z}(v| \vp_n)| \vp_n\ket\langle \vp_n|.
\eeqt
Now $V_v$ can be considered as a possibly unbounded operator on $\hil$ with the domain $\h V$. The operator $V_v$ is bounded
if and only if $\sup_{n\in\Z}|(v|\vp_n)|<\infty$, that is, $v\in\hil_\infty$, in which case we consider it defined on the whole $\hil$.

For each pair $v,u$ of generalized vectors we can construct a matrix $C$ with $c_{nm} = (\vp_n|v)(u|\vp_m)$ for all $n,m\in \Z$.
The associated generalized operator measure $G^C$, which we denote by $G^{v,u}$, can then be written in the form
\bet
G^{v,u}(X)=\frac{1}{2\pi}\int_{X}e^{i\theta Z}|v)(u|e^{-i\theta Z}d\theta,
\eeqt
in the sense that
\bet
G^{v,u}_{\vp,\psi}(X)=\frac{1}{2\pi}\int_{X}( e^{-i\theta Z}\vp|v)(u|e^{-i\theta Z}\psi)d\theta, \ \ \vp,\psi\in \h V.
\eeqt
We say that such a $G^{v,u}$ is a \emph{simple} generalized operator measure. Clearly, $G^{v,u}$ has the form
\bet
G^{v,u}= V_v^*G^{\bf 1}V_u,
\eeqt
meaning that $G^{v,u}(X)(\vp,\psi) = \bra V_v\vp|\overline{G^{\bf1}}(X)V_u\psi\ket$ for all $X$.

It is easy to characterize those simple generalized operator measures which determine an operator measure:

\begin{proposition} \label{supprop} Let $v,u\in \h V^\times$, and suppose that $G^{v,u}\ne 0$. Let $C$ denote the structure matrix of $G^{v,u}$.
The following conditions are equivalent.
\begin{itemize}
\item[(i)] $C$ is a Schur multiplier;
\item[(ii)] $G^{v,u}$ is extensible;
\item[(iii)] $v,u\in\hil_\infty\setminus\{0\}$.
\end{itemize}
\end{proposition}
\begin{proof}
That (i) implies (ii) is clear.

Suppose now that (ii) holds. Clearly $v\neq 0$ and $u\neq 0$.
If e.g. $\sup_{n\in\Z}|(\vp_n|v)|=\infty$ and $m\in\Z$ such that $(\vp_m|u)\ne 0$, we get $\sup_{n\in\Z}|c_{nm}|=\infty$, which contradicts the
assumption according to Proposition \ref{propcyclic}. This proves (iii).

If (iii) holds, then $C$ is a Schur multiplier, since $c_{nm}= (\vp_n|v)(u|\vp_m)$ (e.g. $\vp_n = (v |\vp_n)\vp_0$ and 
$\psi_n = (u |\vp_n)\vp_0$
define bounded sequences of vectors).
\end{proof}

The following result is immediate:

\begin{proposition} Any generalized operator measure can be written in terms of simple generalized operator measures in the form
\bet
G^C=\sum_{k\in\Z}G^{v^k,u^k},
\eeqt
where $v^k,u^k\in \h V^\times$ for all $k\in\Z$. Moreover, if $\sup_{n,m\in\Z}|c_{nm}|<\infty$, then the generalized vectors
$v^k$ and $u^k$ can be chosen so that each $G^{v^k,u^k}$ is extensible.
\end{proposition}
\begin{proof}
Choose, for example, $v^k=\vp_k$ and $u^k=\sum_{m\in\Z}\overline{c_{k,m}}\vp_m$, or
$v^k=\sum_{n\in\Z}c_{n,k}\vp_n$ and $u^k=\vp_k$. If $\sup_{n,m\in\Z}|c_{nm}|<\infty$ then $v^k,u^k\in \hil_\infty$ for all $k$, so
each $G^{v^k,u^k}$ is extensible by Proposition \ref{supprop}.
\end{proof}

\begin{proposition}
Let $C$ be a matrix. Then $C$ is a Schur multiplier if and only if there exists a decomposition
\bet
G^C=\sum_{k\in\Z}G^{v^k,u^k}
\eeqt
such that  $\sup_{n\in\Z}\{\sum_{k\in\Z}|(\vp_n|v^k)|^2\}$ and
$\sup_{n\in\Z}\{\sum_{k\in\Z}|(\vp_n|u^k)|^2\}$ are finite.
\end{proposition}
\begin{proof}
Let $C$ be a Schur multiplier and let $(\psi_n),(\eta_n)$ be bounded sequences in the  Hilbert space $\hil$ such that
$c_{nm} = \bra\psi_n|\eta_m\ket$ for all $n,m\in \Z$. 
For each $k\in\Z$, define the generalized vectors $v^k$ and $u^k$ by $(v^k |\vp_n) =
\bra \vp_k|\psi_n\ket$, and $(u^k| \vp_n) = \bra \vp_k|\eta_n\ket$ for all $k\in\Z$. Now
$c_{nm}=\sum_{k\in\Z}(\vp_n|v^k)(u^k|\vp_m)$,
and hence
$G^C=\sum_{k\in\Z}G^{v^k,u^k}$.
The statement $\sup_{n\in\Z}\{\sum_{k\in\Z}|(\vp_n|v^k)|^2\}<\infty$ and $\sup_{n\in\Z}\{\sum_{k\in\Z}|(\vp_n|u^k)|^2\}<\infty$ are
equivalent to the boundedness of the sequences $(\psi_n)$ and $(\eta_n)$.

On the other hand, if we assume that there exists such a decomposition, then the above formulas define $\psi_n$ and $\eta_n$ in terms
of the generalized vectors $v^k$ and $u^k$.
\end{proof}

Finally, we consider briefly the possibility of reducing the extensibility problem to the positive case. 

Let
$G^C=\sum_{k\in\Z}G^{v^k,u^k}$
be a generalized operator measure.
Using the polarization identity
we can write $G^C$ as a sum of four positive generalized operator measures as follows:
$$
G^C=\frac{1}{4}\sum_{s=0}^3i^s G_s^C
$$
where 
$G_s^C=\sum_{k\in\Z}G^{v^k+i^su^k,v^k+i^su^k}$.
Since $G_s^C$ is positive, it is extensible if and only if its structure matrix is a Schur multiplier, i.e., 
$\sup_{n\in\Z}\sum_{k\in\Z}|(\vp_n|v^k+i^su^k)|^2<\infty.$ If this happens for all $s=0,1,2,3$, then  $C$ itself is a Schur multiplier, 
and $G^C$ is extensible.

\section{On norms of Schur multipliers}\label{normsec}

In this Section we consider four norms of Schur multipliers which arise naturally in the context of covariant operator measures.

As mentioned before, the set $\h S$ of Schur multiplier matrices constitutes a subset of the set
\bet
\h C=\{ C = (c_{nm})\mid G^C \text{ extensible}\}
\eeqt
The set $\h C$ is in a bijective correspondence with the set $\h E_{cov}= \{ E\in \h E\mid E \text{ covariant}\}$, where
$\h E$ is the set of all operator measures ${\bo}\to L(\hil)$. 
Clearly, all the sets $\h S$, $\h C$, $\h E_{cov}$ and $\h E$ are linear spaces, and the bijection $\h C\ni C\mapsto \overline{G^C}\in \h E_{cov}$ is
linear. Hence, we can endow the set $\h C$ with a natural norm of the set $\h E$, which was defined in Section \ref{opsec}. 
In other words, we have a norm $\| \cdot\|_o$ on $\h C$ defined by $\|C\|_o = \|G^C\|$, and the spaces
$\h C$ and $\h E_{cov}$ are isometrically isomorphic. As a subspace of $\h C$, also the space of Schur multipliers now becomes a
normed space.

\begin{proposition} The space $(\h C, \|\cdot\|_o)$ is Banach.
\end{proposition}
\begin{proof} The space $\h E$ is a subspace of the set $\h B$ of all bounded functions $f:{\bo}\to L(\hil)$, which is a Banach space
with respect to the uniform norm (because $L(\hil)$ is Banach). Clearly, the norm on $\h E_{cov}$ is precisely the uniform norm, so it suffices to show that
$\h E_{cov}$ is a closed subspace of $\h B$. To that end, let $(E_n)$ be a sequence of covariant operator measures converging
to a bounded map $f:{\bo}\to L(\hil)$ uniformly in ${\bo}$. It is clear that $f$ is then additive. Since the convergence is
uniform, it follows by a standard $\frac{\epsilon}{2}$-argument that for each fixed $\vp,\psi\in \hil$, the map
$X\mapsto \bra \vp |f(X)\psi\ket$ is complex measure.
Hence, $f\in \h E$. Moreover, because
\bet
e^{i\theta Z}E_n(X)e^{-i\theta Z}= E_n(X+\theta\,({\rm mod} 2\pi)), \ \ \  X\in \hB([0,2\pi)), \theta\in [0,2\pi),
\eeqt
and $(E_n(X))$ converges in norm to $f(X)$ for all $X$, it follows that also $f$ is covariant, i.e.
$f\in \h E_{cov}$. Hence, $\h E_{cov}$ is closed. The proof is complete.
\end{proof}

 The most natural norm of a Schur multiplier $S$ is of course the one arising directly from its definition, namely
the norm $\|\cdot\|_m$ of the Schur product map $L(\hil) \ni A \mapsto S* A\in L(\hil)$ (see e.g. \cite[p. 108]{Paulsen}).
We note that $\|S\|_o\leq \|S\|_m$ for each Schur multiplier $S$. 
Indeed
\bet
\|S\|_o = \sup_{X\in{\bo}} \|\overline{G^S}(X)\| = \sup_{X\in{\bo}} \|S* i(X)\|\leq \sup_{A\in L(\hil), \|A\|\leq 1} \|S* A\| = \|S\|_m.
\eeqt
However, these norms are not the same. Indeed, let $S=(s_{nm})$ be a matrix with $s_{01}=1$ and all the other entries zero. For
a matrix $A=(a_{nm})$, the norm of the matrix $S* A$ is clearly $|a_{01}|$. Hence, for $\|A\|\leq 1$,
we have $\|S* A\| = |a_{01}|\leq 1$ (because $|a_{01}|^2\leq \sum_{k} |a_{k1}|^2\leq \|A\|^2\leq 1$). In addition, $\|S* S\|=1$,
so $\|S\|_m =1$. But $\|S* i(X)\| = |i_{01}(X)|\leq \frac {\sqrt{2}}{\pi}$ for all $X\in \bo$. (This is because
\bet
\left(\frac{1}{2\pi}\right)^{-2}|i_{01}(X)|^2= \left|\int_X e^{-i\theta} \ d\theta\right|^2 = \left|\int_X \cos \theta \ d\theta\right|^2+\left|\int_X \sin \theta \ d\theta\right|^2
\leq 4+4 = 8,
\eeqt
where the last inequality is obtained by using the estimate
\bet
\left|\int_X \sin\theta \ d\theta\right|
\leq \max \left\{ \int_{X\cap [0,\pi)} \sin\theta \ d\theta, \int_{X\cap [\pi,2\pi)} -\sin\theta \ d\theta\right\} \leq \int_0^{\pi} \sin \theta \ d\theta = 2,
\eeqt
and the corresponding one involving $\cos \theta$.)
Hence, $\|S\|_o \leq \frac {\sqrt{2}}{\pi} <1 =\|S\|_m$, so the norms are different. Of course, it could still be that for some constant $M>0$,
we had $\|\cdot \|_m= M\|\cdot \|_o$. This possibility is ruled out by the example where $S'=(s'_{nm})$, with $s_{00}=1$, and the other
entries zero. Namely, then we have $\|S\|_m=1$, as before, but now also $\|S\|_o=1$, since $i_{00}([0,2\pi)) = 1$. Hence, the norms are
not constant multipliers of each other.

The question remains of whether these norms are equivalent or not (i.e.\ do they determine the same topology).

\

For any element $C\in\h C$ it is necessary that 
\begin{enumerate}
\item $\|C\|_{1,\infty}=S^C<\infty$,
\item the first moment form $\Theta_1^C$ is bounded.
\end{enumerate}
Therefore,
we can define two natural norms, namely, $C\mapsto\|C\|_{1,\infty}$ and $C\mapsto\|C\|_f=\|\Theta_1^C\|$ .
It is easy to see that the first mapping is actually a norm and,
by the proof of Proposition \ref{propcyclic}, we have $\|C\|_{1,\infty}\leq 4 \|C\|_o$ for each
$C\in\h C$. Since $C\mapsto\Theta_1^C$ is linear and $\|\Theta_1^C\|=0$ implies that $C$ is a zero matrix, the mapping $C\mapsto\|C\|_f$ is a norm also.

\

An important subset of $\h C$ is the set $\h C_+^1$ of matrices which determine covariant observables. 
Following \cite{PellonpääIII} it can be shown that $\h C_+^1$ consists of positive semidefinite matrices with unit diagonals. Hence, $\h C_+^1$ is also a subset of the set $\h S$ of Schur multipliers. Moreover, $\h C_+^1$ is closed under the Schur multiplication. The following observation shows that $\h C_+^1$ is contained in the surface of the closed unit ball, with respect to the first three norms considered above.

\begin{proposition} $\|C\|_{1,\infty}=\|C\|_m=\|C\|_o=1$ for all $C\in{\h C}_+^1$.
\end{proposition}
\begin{proof}
Let $C\in{\h C}_+^1$. Since $C$ is positive semidefinite with the unit diagonal, we have $|c_{nm}|\le 1$ and $|c_{nn}|=1$, and so it follows that $\|C\|_{1,\infty}=1$. Moreover, there exists a sequence $(\psi_m)$ of unit vectors such that $c_{nm}=\bra\psi_n|\psi_m\ket$ for all $n,m$, so one gets $\|C\|_m\le1$ by \cite[Corollary 8.8]{Paulsen}. In addition, $\|C*I\|=\|I\|=1$, so that $\|C\|_m=1$. Finally, since  for all $X\in\bo$, $O\le\overline{G^C}(X)\le \overline{G^C}([0,2\pi))=I$ and, thus, $\|\overline{G^C}(X)\|\le \|\overline{G^C}([0,2\pi))\|=1$, it follows that $\|C\|_o=1$. 
\end{proof}

Defining a unitary transform $U_\upsilon=\sum_{n\in\Z}e^{i\upsilon_n}|n\ket\bra n|$, $(\upsilon_n)\subset [0,2\pi)$ (which commutes with $Z$), for any $C\in\h C_+^1$, one gets a family of (physically) equivalent covariant observables 
$$
\big\{U_\upsilon\overline{ G^C}U_\upsilon^*\,\big|\,(\upsilon_n)\subset [0,2\pi)\big\}.
$$
The structure matrix of $U_\upsilon\overline{ G^C}U_\upsilon^*$ is $C*Y_\upsilon$ where 
$Y_\upsilon=(e^{i(\upsilon_n-\upsilon_m)})_{n,m\in\Z}\in\h C_+^1$.
Thus, it is reasonable to define an equivalence relation on $\h C$ (and thus on $\h C_+^1$) by declaring any two matrices $C$ and $D$ equivalent if $D=C*Y_{\upsilon}$
for some $Y_v$. 
The set ${\bf C}_+^1$ of equivalence classes of $\h C_+^1$  can then be equipped with the partial ordering
$[C]\preccurlyeq[D]$ if $C=D*E$ for some $E\in \h C_+^1$ \cite{PellonpääIII}.
Since $C={\bf 1}*C$, we have $[C]\preccurlyeq[\bf 1]$ for all $\h C_+^1$, so $[\bf 1]$ 
is the upper bound in ${\bf C}_+^1$; similarly, since $I=C*I$, we get $[I]\preccurlyeq [C]$ and hence $[I]$ is the lower bound in ${\bf C}_+^1$.

Define then a mapping 
$$
\alpha:\,{\bf C}_+^1\to[0,\infty),\;[C]\mapsto\|C\|_f,
$$
which is well-defined since $\|C*Y_\upsilon\|_f=\|U_\upsilon\Theta_1^CU_\upsilon^*\|=\|C\|_f$.
\begin{proposition}
The mapping $\alpha$ from $({\bf C}_+^1,\preccurlyeq)$ to $([\pi,2\pi],\le)$ is order preserving, i.e.\
$$ 
\pi=\alpha([I])\le\alpha([C])\le\alpha([D])\le\alpha([{\bf 1}])=2\pi
$$
for all $[C],\,[D]\in{\bf C}_+^1$ such that
$[C]\preccurlyeq[D]$.
\end{proposition}

\begin{proof}
Let then $C,\,D\in\h C_+^1$ be such that 
$[C]\preccurlyeq[D]$, i.e., $C=D*E$ for some $E\in \h C_+^1$.
It follows that $\Theta_1^C=E*\Theta_1^D$ and, since $E$ is, as a Schur multiplier, continuous,
$\|\Theta_1^C\|\le\|E\|_m\|\Theta_1^D\|=\|\Theta_1^D\|$. Moreover, $\|I\|_f=\|\pi I\|=\pi$ and,
as a multiplication operator on $L^2([0,2\pi))$, $\|{\bf 1}\|_f=\|\Theta_1^{\bf 1}\|=2\pi$.
\end{proof}

To conclude, the norms $\| \, \cdot  \, \|_{1,\infty}$, $\|\,\cdot \, \|_m$, and $\|\,\cdot \, \|_o$
cannot distinguish the elements of $\h C_+^1$, but
$\|\,\cdot \, \|_f$ defines an order preserving mapping $\alpha$, which gives an (intuitively) correct ordering to the set of covariant observables.

\section{Summary}
The main results of this paper are summarized as follows.

\noindent
{\bf Boundedness conditions:}
\begin{itemize}
\item[(a)] If $C$ is a matrix such that the generalized operator measure $G^C$ is extensible, then the following
boundedness conditions hold.
\begin{enumerate}
\item $S^C<\infty$;
\item The first moment of $G^C$ is bounded.
\end{enumerate}
\item[(b)] Neither (1) nor (2) implies the extensibility of $G^C$.
\end{itemize}

\noindent
{\bf Implications of the boundedness conditions:}
Assume that the boundedness conditions (1) and (2) hold. Then
\begin{itemize}
\item[(i)] All cyclic moment forms are bounded (with the same constant).
\item[(ii)] All moment forms are (exponentially) bounded.
\item[(iii)] $G^C([a,b))\in L(\hil)$ for all $0\le a\le b\le 2\pi$.
\item[(iv)] $\int f d G^C\in L(\hil_1,\hil)$ for all bounded $f:[0,2\pi)\to \C$; especially,
$G^C(X)\in L(\hil_1,\hil)$ for all $X\in\bo$.
\item[(v)] For any $\vp\in\hil$, $\psi\in\hil_1$, there exists a complex measure, which
coincides with $G^C_{\vp,\psi}$ whenever $\vp,\psi\in \h V$. This measure has an $L^2$-density.
\end{itemize}

\noindent
{\bf Schur multipliers:}
\begin{itemize}
\item[(a)] If $C$ is a Schur multiplier, then $G^C$ is extensible.
\item[(b)] If $C$ is positive semidefinite, then it is a Schur multiplier if and only if $G^C$ is extensible.
\item[(c)] If $G^C$ is a simple generalized operator measure, then it is a Schur multiplier if and only if $G^C$ is extensible.
\item[(d)] Each $G^C$ can be written as a (weak) sum of simple generalized operator measures.
\end{itemize}

Finally, we observed that, in addition to the well-known norms $\|\cdot\|_m$ and $\| \cdot \|_{1,\infty}$ of Schur multipliers, we have two other  norms
$\|\cdot\|_o$ and $\|\cdot\|_f$ for them, arising from the connection to operator measures.
The set of covariant observables is contained in the surface of unit ball with respect to the norm $\|\cdot\|_o$, while the norm $\|\cdot\|_f$ turns out to be compatible with the
(partial) ordering of covariant observables.

\

\noindent{\bf Acknowledgments.} The authors thank Prof. Kari Ylinen for fruitful discussions. One of us (J.K.) was supported by Turku University Foundation.

\end{document}